\documentstyle[12pt]{article}
\begin{document}
\hbadness=10000
\hbadness=10000
\begin{titlepage}
\nopagebreak
\begin{flushright}
{\normalsize
HIP-1998-16/TH\\
March, 1998}\\
\end{flushright}
\vspace{0.7cm}
\begin{center}

{\large \bf  Softly Broken $N=1$ Supersymmetric QCD}

\vspace{1cm}

{\bf Masud Chaichian  
\footnote[1]{e-mail: chaichia@rock.helsinki.fi}}, 
{\bf Wen-Feng Chen  
\footnote[2]{e-mail: wfchen@rock.helsinki.fi}}\\ 
and\\
{\bf Tatsuo Kobayashi 
\footnote[3]{e-mail: kobayash@rock.helsinki.fi}}\\

\vspace{0.7cm}
Department of Physics, High Energy Physics Division\\
    University of Helsinki \\
and\\
Helsinki Institute of Physics\\
    P.O. Box 9 (Siltavuorenpenger 20 C)\\
    FIN-00014 Helsinki, Finland \\

\end{center}
\vspace{0.7cm}

\nopagebreak

\begin{abstract}
We study softly broken $N=1$ supersymmetric QCD with the gauge group 
$SU(N_c)$ and $N_f$ flavours of quarks for $N_f > N_c+1$.
We investigate the phase structure of its dual theory adding generic 
soft supersymmetry breking terms, i.e. soft scalar masses, 
trilinear coupling terms of scalar fields  and gaugino masses.
It is found that the trilinear coupling terms play an improtant role 
in determining the potential minima.
Also we compare softly broken original and dual theories in the broken phase.
\end{abstract}
\vfill
\end{titlepage}
\pagestyle{plain}
\newpage
\def\thefootnote{\fnsymbol{footnote}}

\section{Introduction}
In the last few years, the understanding of strongly coupled 
supersymmetric (SUSY) Yang-Mills theories has made important progress.
A number of exact results have been obtained \cite{rev}, 
because holomorphy and 
global symmetries including $R$-symmetry restrict the possible structure of 
the theory \cite{seiberg1}.
In paricular, the old Montonen-Olive duality conjecture \cite{MO}
has been gained new insights, although this duality 
was shown to possibly exist only in $N=4$ SUSY 
Yang-Mills theory\cite{osborn}.
Seiberg and Witten have found the exact low-energy effective action of 
$N=2$ SUSY gauge field theory and its vacuum structure through 
duality \cite{SW}.
Furthermore, Seiberg's duality was found to be realized in the 
infrared region of $N=1$ SUSY QCD with the $SU(N_c)$ 
gauge group and $N_f$ flavours of quarks for the case  
$N_f > N_c +1$ \cite{seiberg2}, which actually includes QCD in 
the real world, i.e. $(N_c,N_f)=(3,6)$.
This conjecture is strongly supported by the 't Hooft anomaly 
matching condition \cite{thooft}.
Since duality can relate a strongly coupled theory 
to a weakly coupled one, thus this has provided with a possible way 
to explore the non-perturbative aspects of the strongly coupled theory.

It is important to extend such analyses to non-SUSY cases.
However, it is very difficult to discuss strongly coupled 
non-SUSY Yang-Mills theories directly, since 
we can not use holomorphy or $R$-symmetry in a non-SUSY 
thoery.
Thus, for the study of a non-SUSY theory it is of much interest  
to consider $N=1$ SUSY theory with soft SUSY breaking terms.
Actually in Ref.[8-10] $N=1$ SUSY QCD 
with soft scalar masses 
as well as with gaugino masses was discussed and interesting 
results were obtained \footnote{Softly broken $N=2$ SUSY QCD 
has been studied in \cite{soft4}.}.

If all the symmetries allow it, then other soft SUSY breaking terms, i.e. 
trilinear ($A$-terms) and bilinear ($B$-terms) interactions of 
scalar fields in general also appear in the low energy 
effective theory, e.g. from the viewpoint of 
supergravity or dynamical SUSY breaking.
These terms are very important to determine the potential minima.
For instance, in the minimal supersymmetric standard model, 
the successful electroweak symmetry breaking can not be realized without 
the $B$-term of two Higgs fields \cite{Higgs} and large values of 
$A$-terms lead to charge and/or colour breaking vacua \cite{CCB}.

When the superpotential $W$ includes the term 
$\widehat q \widehat T \widehat {\bar q}$, the 
scalar potential  in general also has  
the corresponding $A$-term $qT\bar q$, where $q$ ($T, \bar q$) denotes 
the scalar component corresponding to the supermultiplet 
$\widehat q$ ($\widehat T, \widehat {\bar q}$).
The above superpotential corresponds to the dual to 
$N=1$ SUSY QCD theory with the gauge group $SU(N_c)$ and $N_f$ flavours of 
quark supermultiplets $\widehat Q$ and 
$\widehat {\bar Q}$ for $N_f > N_c+1$ \cite{seiberg2}.
In this case $\widehat q$ and $\widehat {\bar q}$ are the dual quark 
superfields and $\widehat T$ is the meson superfield.
In this paper we study such a dual theory with soft masses and 
$A$-terms.
We investigate the phase structure of this theory taking the 
soft parameters as free parameters.
If the fundamental theory for SUSY breaking is specified, e.g. originated 
from  certain type of supergravity model, 
superstring theory \cite{ST-soft} 
or dynamical SUSY breaking \cite{dsb} \footnote{Application 
of coupling reduction theory \cite{cred} to the  
soft SUSY breaking terms is another type of interesting approach to 
fix the relations among soft SUSY breaking terms \cite{RGsoft}.}, 
these soft parameters can be written in terms of more fundamental 
quantities such as $F$-term condensations.
However, here it is more instructive to take the soft parameters 
as free parameters 
in order to understand the generic phase structure of the theory.
We also compare softly broken dual theory with 
softly broken original one.

This paper is organized as follows.
In section 2 we give a brief review of $N=1$ SUSY QCD and its Seiberg's 
dual theory.
We add the generic soft SUSY breaking terms to the dual theory and 
study its phase structure.
In section 3 we compare this phase structure of the dual theory with 
softly broken original $N=1$ SUSY QCD theory.
Section 4 is devoted to conclusions and discussions.

\section{Softly broken dual to $N=1$ SUSY QCD theory}
At first we review briefly $N=1$ SUSY QCD and 
its dual theory \cite{seiberg2}.
Here we consider $N=1$ SUSY QCD with the gauge group $SU(N_c)$ 
and $N_f$ flavours of quark supermultiplets 
$\widehat Q^i$ and $\widehat {\bar Q}_i$ 
($i=1,\cdots , N_f$), 
where $\widehat Q$ and $\widehat {\bar Q}$ transforms 
under $N_c$ and $\bar N_c$ representations of $SU(N_c)$.
Hereafter the colour indices are omitted.
This theory has a vanishing superpotential and has the global symmetry 
\begin{equation}
SU(N_f)_{Q}\times SU(N_f)_{\bar Q} \times U(1)_B \times U(1)_R.
\end{equation}
Quark superfields $\widehat Q$ and $\widehat {\bar Q}$ transform as    
the multiplets $(N_f,0,1)$ and $(0,\bar N_f,1)$ of the global symmetry 
$SU(N_f)_{Q}\times SU(N_f)_{\bar Q} \times U(1)_B$, respectively.

Its dual theory has the gauge group $SU(\tilde N_c)$, 
where $\tilde N_c \equiv N_f-N_c$, and contains $N_f$ flavours of 
dual quark supermultiplets 
$\widehat q_i$ and $\widehat {\bar q}^i$ and 
singlet superfields $\widehat T^i_j$, 
which correspond to meson supermultiplets in the original theory.
This dual theory has the same global symmetry as the original one and 
the superfields $\widehat q_i$, $\widehat {\bar q}^i$ and $\widehat T^i_j$ 
transform as  $(\bar N_f,0,N_c/\tilde N_c)$, 
$(0,N_f,-N_c/\tilde N_c)$ and $(N_f,\bar N_f,0)$ 
under the global symmetry 
$SU(N_f)_{q}\times SU(N_f)_{\bar q} \times U(1)_B$, respectively.
The dual theory has the superpotential,
\begin{equation}
W=\widehat q_i \widehat T^i_j \widehat {\bar q}^j.
\end{equation}
This dual pair has the same anomaly structure for the global symmetries, 
i.e. the 't Hooft anomaly matching condition \cite{thooft} is satisfied.
The 't Hooft anomaly matching condition plays basic role in the probe of 
dual pairs, i.e. massless fermions and their global symmetries 
are important.

Now let us consider the non-SUSY case.
Here we break $N=1$ SUSY softly.
We add the following soft SUSY breaking terms to the dual theory:
\begin{eqnarray}
{\cal L}_{SB} &=& -m_q^2 {\rm tr} |q|^2-m_{\bar q}^2 {\rm tr} |\bar q|^2 - 
m_T^2 {\rm tr} |T|^2+(hq_iT^i_j\bar q^j + h.c.).
\end{eqnarray}
Also the gaugino mass terms are added.
Here soft scalar mass terms and the $A$-term are flavour-independent.
Note that these terms are all the possible soft terms to be added and 
they do not break any global symmetry 
except $R$-symmetry.
For the kinetic term, we assume the canonical form with 
normalization factors $k_q$ and $k_T$ for $q$, $\bar q$ and $T$.
Then we write the following scalar potential:
\begin{eqnarray}
V(q,\bar q,T) &=& {1 \over k_T}{\rm tr} (q q^\dagger \bar q^\dagger \bar q)
+{1 \over k_q}{\rm tr}(qTT^\dagger q^\dagger
+\bar q^\dagger T^\dagger T \bar q) \nonumber \\
&+& {\tilde g^2 \over 2}({\rm tr} q^\dagger \tilde t^a q -  
{\rm tr} \bar q \tilde t^a \bar q^\dagger)^2 
+m_q^2{\rm tr} q^\dagger q + m_{\bar q}^2 {\rm tr} \bar q \bar q^\dagger  
\nonumber \\ 
&+& m_T^2 {\rm tr} T^\dagger T - (hq_iT^i_j\bar q^j + h.c.),
\end{eqnarray}
where the third term is the $D$-term and $\tilde g$ denotes 
the gauge coupling constant of the dual theory.

We assume $h$ is real.
The minimum of potential can be obtained along the 
following diagonal direction \cite{soft2},
\begin{equation}
q=\left( \begin{array}{ccccc}
q_{(1)} &  & & 0 & \\
 &  q_{(2)} & & & \\
 0 &  & \cdots &  & \\
 & & &   q_{(\tilde N_c)} & 
   \end{array}\right),
\end{equation}
\begin{equation}
\bar q=\left( \begin{array}{cccc}
\bar q_{(1)} &  & & 0 \\
 &  \bar q_{(2)} & & \\
0 &  & \cdots &  \\
 & & &   \bar q_{(\tilde N_c)} \\
 & & & 
   \end{array}\right),
\end{equation}
\begin{equation}
T=\left( \begin{array}{ccccc}
T_{(1)} & & &  0 & \\
 &  T_{(2)} & & & \\
 0 &  & \cdots &  & \\
 & & & T_{(\tilde N_c)} & \\
 & & & & 0 
   \end{array}\right),
\end{equation}
where all the entries, $q_{(i)}$, $\bar q_{(i)}$ and $T_{(i)}$, 
can be made real.
In this case the scalar potential is written as 
\begin{eqnarray}
V(q,\bar q,T) &=&
{1 \over k_T}\sum^{\tilde N_c}_{i=1} q^2_{(i)} \bar q^2_{(i)} 
+{\tilde g^2 \over 4 \tilde N_c}\sum^{\tilde N_c}_{i<j}
(q^2_{(i)}-\bar q^2_{(i)} - q^2_{(j)}+\bar q^2_{(j)})^2 \nonumber \\
&+& m_q^2 \sum^{\tilde N_c}_{i=1} q^2_{(i)} 
+m_{\bar q}^2 \sum^{\tilde N_c}_{i=1} \bar q^2_{(i)} 
+m_T^2 \sum^{\tilde N_c}_{i=1} T^2_{(i)} \nonumber \\
&+& {1 \over k_q}\sum^{\tilde N_c}_{i=1} 
T^2_{(i)}(q^2_{(i)}+\bar q^2_{(i)}) -2h \sum^{\tilde N_c}_{i=1} 
q_{(i)} T_{(i)} \bar q_{(i)}.
\label{pot1}
\end{eqnarray}

Let us study the minimum of the potential (\ref{pot1}).
For fixed values of $q_{(i)}$ and $\bar q_{(i)}$, this potential 
is unbounded from below along $T \rightarrow \infty$, if 
$m^2_T+(q_{(i)}+\bar q_{(i)})/k_q < 0$, which corresponds to 
$m^2_T<0$ in the limit $q_{(i)}=\bar q_{(i)}=0$.

The stationary condition, $\partial V/\partial T_{(i)}=0$, requires 
\begin{equation}
T_{(i) min}={h q_{(i)}q_{(i)} \over m^2_T + (q_{(i)}+\bar q_{(i)})/k_q }.
\label{Tmin}
\end{equation}
Using this, we write the scalar potential as 
\begin{eqnarray}
V(q,\bar q,T_{min}) &=&
{1 \over k_T}\sum^{\tilde N_c}_{i=1} q^2_{(i)} \bar q^2_{(i)} 
+{\tilde g^2 \over 4 \tilde N_c}\sum^{\tilde N_c}_{i<j}
(q^2_{(i)}-\bar q^2_{(i)} - q^2_{(j)}+\bar q^2_{(j)})^2 \nonumber \\
&+& m_q^2 \sum^{\tilde N_c}_{i=1} q^2_{(i)} 
+m_{\bar q}^2 \sum^{\tilde N_c}_{i=1} \bar q^2_{(i)} \nonumber \\ 
&-&  
\sum^{\tilde N_c}_{i=1} {h^2 q^2_{(i)} \bar q^2_{(i)} \over 
m^2_T + (q_{(i)}+\bar q_{(i)})/k_q }.
\end{eqnarray}
Quartic terms appear in the first and second terms as 
well as  the last term.
The potential minimum corresponds to the direction along which some of 
these quartic terms vanish.
Note that if the first term vanishes, the last term also vanishes.

Let us study the direction where 
\begin{equation}
q_{(i)}=q, \quad \bar q_{(i)}=0.
\end{equation}
In this direction each quartic term disappears and $T_{(i) min}$ also 
vanish.
In this case we have the scalar potential as $V=\tilde N_c m_q^2 q^2$.
Thus, the scalar potential is unbounded from below along the direction 
$q \rightarrow \infty$, if $m_q^2 <0$.
If $m_q^2 >0$, the scalar potential has the minimum $V=0$ at $q=0$.
For the  direction with $q_{(i)}=0$ and $\bar q_{(i)}=\bar q$, we have 
similar results, i.e. the potential is unbounded from below if 
$m_{\bar q}^2<0$.

The second term in (\ref{pot1}), i.e. the $D$-term, vanishes 
along the following direction:
\begin{equation}
q_{(i)}=\bar q_{(i)}=X_i.
\end{equation}
Along this direction, the scalar potential (\ref{pot1}) is written as 
\begin{eqnarray}
V(X,T) &=& \sum^{\tilde N_c}_{i=1}[{1 \over k_T}X_i^4+
(m_q^2+m_{\bar q}^2)X_i^2+m_T^2T_{(i)}^2 \nonumber \\
&+& {2 \over k_q}T_{(i)}^2X_i^2 - 2hT_{(i)}X_i^2].
\end{eqnarray}
Note that the $i$-th elements, i.e. $X_i$ and $T_{(i)}$, are decoupled 
from the $j$-th elements ($i \neq j$).
The stationary condition, $\partial V /\partial X_i=0$, requires 
\begin{equation}
{2 \over k_T}X_i^3+[(m_q^2+m_{\bar q}^2)
+{2 \over k_q}T_{(i)}^2 - 2hT_{(i)}]X_i=0.
\label{stat2}
\end{equation}
If this equation as well as eq.(\ref{Tmin}) has a solution except $X_i =0$, 
this point then corresponds to the potential minimum 
which has a lower energy than $V=0$ given at the origin 
$X_i=T_{(i)}=0$.
Recall that if $X_i =0$, $T_{(i)}$ always vanishes owing to eq.(\ref{Tmin}).
One of the conditions leading to the broken phase, i.e. 
$X_i \neq 0$, is obtained from 
\begin{equation}
f(T_{(i)}) \equiv {2 \over k_q}T_{(i)}^2 - 2hT_{(i)}
+m_q^2+m_{\bar q}^2 \leq 0.
\end{equation}
Otherwise, we always have the unbroken phase, i.e. $X_i=T_{(i)}=0$.
The above inequality is satisfied for the values of $T_{(i)}$
\begin{equation}
{h-\sqrt {h^2-2(m_q^2+m_{\bar q}^2)/k_q } \over 2/k_q} \leq T_{(i)} 
\leq {h+\sqrt {h^2-2(m_q^2+m_{\bar q}^2)/k_q } \over 2/k_q },
\label{region}
\end{equation}
if 
\begin{equation}
h^2 \geq {2 \over k_q}(m_q^2+m_{\bar q}^2).
\label{cond1}
\end{equation}
The inequality (\ref{cond1}) is one of conditions on 
soft SUSY breaking parameters to realize the broken phase.
Furthermore, both of stationary conditions, 
$\partial V/T_{(i)} = \partial V/X_i =0$, should be satisfied.
That leads to the following trilinear equation for $T_{(i)}$ through 
the use of eqs. (\ref{Tmin}) and (\ref{stat2}),
\begin{equation}
g(T_{(i)}) \equiv 
({2 \over k_q}T_{(i)}-h)f(T_{(i)})-{2 \over k_T}m_T^2T_{(i)}=0.
\end{equation}
If this equation has a solution in the region given by (\ref{region}), 
the broken phase is realized.
If the inequality (\ref{cond1}) is satisfied and 
$m_T^2$ is not negative, the function $g(T_{(i)})$ 
has always a local maximum point.

Suppose, for a while, that $h > 0$.
Note that the values of $g(T_{(i)})$ at the boundaries 
of the region (\ref{region}) are negative if $m_T^2 > 0$.
Thus, it is the condition for the broken phase that  
the local maximun point of $g(T_{(i)})$ should be within the 
region (\ref{region}) and at that point the value of $g(T_{(i)})$ 
should not be negative.
That leads to the following condition 
\begin{equation}
[{1 \over 3}h^2+{2m_T^2 \over 3k_T}-{4 \over 3}
m_{av}^2)]^{3} - \left( {m_T^2 \over k_T}h \right)^2 
\geq 0,
\label{cond2}
\end{equation}
where $m_{av}^2=(m_q^2+m_{\bar q}^2)/(2k_q)$.
Similarly we can calculate the case with $h < 0$.
The above condition (\ref{cond2}) is available for both cases, $h> 0$ 
and $h < 0$.
Here we define the ratio $\rho$,
\begin{equation}
\rho \equiv {m_T^2 \over m_{av}^2k_T}.
\label{rati}
\end{equation} 
For $\rho \geq 1$, the condition (\ref{cond2}) implies 
\begin{equation}
{h^2 \over m_{av}^2} \geq A_1(\rho), \quad 
A_2(\rho) \geq {h^2 \over m_{av}^2} \geq 
A_3(\rho),
\label{cond21}
\end{equation}
where the points, $h^2=A_i(\rho)m_{av}^2$ for $i=1,2,3$, are the 
boundary points for the inenquialities (\ref{cond2}) to 
become equalities: $A_1(\rho) > A_2(\rho) > A_3(\rho)$.
Fig. 1 shows $A_1(\rho)$ and $A_2(\rho)$ as a function of $\rho$, 
while $A_3(\rho)$ is always negative.
For $\rho < 1$, only the first inequality in (\ref{cond21}) 
is meaningful.
Note that $A_1(\rho) > 4$, for any value of $\rho$.
Then we obtain the broken phase in the soft parameter region 
satisying the conditions (\ref{cond1}) and (\ref{cond2}).
In the region with $A_2(\rho) < 4$, the broken phase is realized 
only for $h^2 \geq A_1(\rho)m_{av}^2 $.
On the other hand, in the region with $A_2(\rho) > 4$ the broken phase 
is realized for $ A_2(\rho)m_{av}^2 \geq h^2 \geq 4 m_{av}^2$, 
as well as for $h^2 \geq A_1(\rho)m_{av}^2 $.
The scalar potential also has the unbouded-from-below directions  
for negative soft scalar mass squared, e.g. for $m_q^2$, $m_{\bar q}^2$ 
and $m_T^2$.
Otherwise, we have the unbroken phase.
Figs. 2 and 3 show this phase structure, as an example  
for $\rho =1$ and 20, respectively.
\newpage
\begin{center}
\input{fig1.tex}

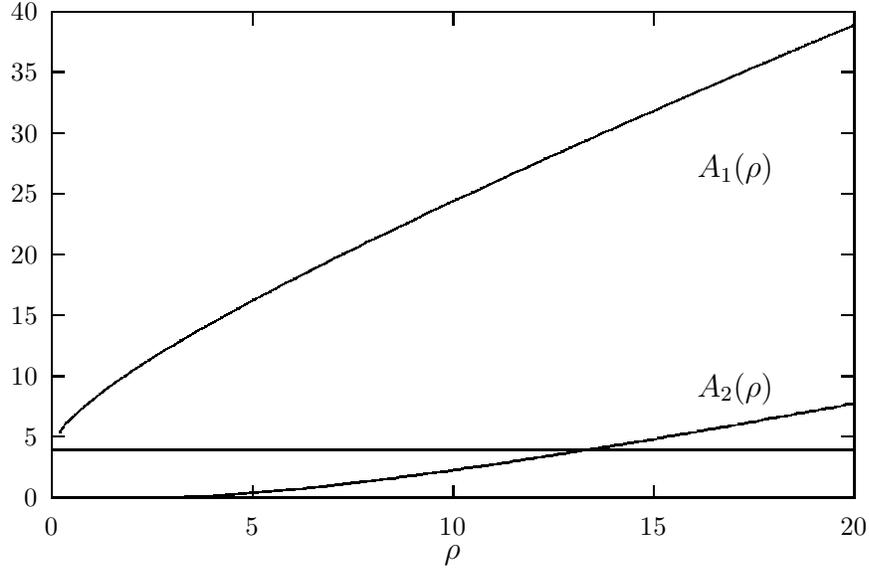
Fig. 1: $A_1(\rho)$ and $A_2(\rho)$ in (\ref{cond21}) as functions of 
$\rho$ defined in (\ref{rati}).
\end{center}
\begin{center}
\setlength{\unitlength}{0.240900pt}
\ifx\plotpoint\undefined\newsavebox{\plotpoint}\fi
\sbox{\plotpoint}{\rule[-0.200pt]{0.400pt}{0.400pt}}%
\begin{picture}(1500,900)(0,0)
\font\gnuplot=cmr10 at 10pt
\gnuplot
\sbox{\plotpoint}{\rule[-0.200pt]{0.400pt}{0.400pt}}%
\put(220.0,113.0){\rule[-0.200pt]{292.934pt}{0.400pt}}
\put(625.0,113.0){\rule[-0.200pt]{0.400pt}{184.048pt}}
\put(220.0,113.0){\rule[-0.200pt]{4.818pt}{0.400pt}}
\put(198,113){\makebox(0,0)[r]{0}}
\put(1416.0,113.0){\rule[-0.200pt]{4.818pt}{0.400pt}}
\put(220.0,232.0){\rule[-0.200pt]{4.818pt}{0.400pt}}
\put(198,232){\makebox(0,0)[r]{5}}
\put(1416.0,232.0){\rule[-0.200pt]{4.818pt}{0.400pt}}
\put(220.0,352.0){\rule[-0.200pt]{4.818pt}{0.400pt}}
\put(198,352){\makebox(0,0)[r]{10}}
\put(1416.0,352.0){\rule[-0.200pt]{4.818pt}{0.400pt}}
\put(220.0,471.0){\rule[-0.200pt]{4.818pt}{0.400pt}}
\put(198,471){\makebox(0,0)[r]{15}}
\put(1416.0,471.0){\rule[-0.200pt]{4.818pt}{0.400pt}}
\put(220.0,591.0){\rule[-0.200pt]{4.818pt}{0.400pt}}
\put(198,591){\makebox(0,0)[r]{20}}
\put(1416.0,591.0){\rule[-0.200pt]{4.818pt}{0.400pt}}
\put(220.0,710.0){\rule[-0.200pt]{4.818pt}{0.400pt}}
\put(198,710){\makebox(0,0)[r]{25}}
\put(1416.0,710.0){\rule[-0.200pt]{4.818pt}{0.400pt}}
\put(220.0,829.0){\rule[-0.200pt]{4.818pt}{0.400pt}}
\put(198,829){\makebox(0,0)[r]{30}}
\put(1416.0,829.0){\rule[-0.200pt]{4.818pt}{0.400pt}}
\put(220.0,113.0){\rule[-0.200pt]{0.400pt}{4.818pt}}
\put(220,68){\makebox(0,0){-2}}
\put(220.0,857.0){\rule[-0.200pt]{0.400pt}{4.818pt}}
\put(423.0,113.0){\rule[-0.200pt]{0.400pt}{4.818pt}}
\put(423,68){\makebox(0,0){-1}}
\put(423.0,857.0){\rule[-0.200pt]{0.400pt}{4.818pt}}
\put(625.0,113.0){\rule[-0.200pt]{0.400pt}{4.818pt}}
\put(625,68){\makebox(0,0){0}}
\put(625.0,857.0){\rule[-0.200pt]{0.400pt}{4.818pt}}
\put(828.0,113.0){\rule[-0.200pt]{0.400pt}{4.818pt}}
\put(828,68){\makebox(0,0){1}}
\put(828.0,857.0){\rule[-0.200pt]{0.400pt}{4.818pt}}
\put(1031.0,113.0){\rule[-0.200pt]{0.400pt}{4.818pt}}
\put(1031,68){\makebox(0,0){2}}
\put(1031.0,857.0){\rule[-0.200pt]{0.400pt}{4.818pt}}
\put(1233.0,113.0){\rule[-0.200pt]{0.400pt}{4.818pt}}
\put(1233,68){\makebox(0,0){3}}
\put(1233.0,857.0){\rule[-0.200pt]{0.400pt}{4.818pt}}
\put(1436.0,113.0){\rule[-0.200pt]{0.400pt}{4.818pt}}
\put(1436,68){\makebox(0,0){4}}
\put(1436.0,857.0){\rule[-0.200pt]{0.400pt}{4.818pt}}
\put(220.0,113.0){\rule[-0.200pt]{292.934pt}{0.400pt}}
\put(1436.0,113.0){\rule[-0.200pt]{0.400pt}{184.048pt}}
\put(220.0,877.0){\rule[-0.200pt]{292.934pt}{0.400pt}}
\put(45,495){\makebox(0,0){$h^2$}}
\put(828,23){\makebox(0,0){$m_{av}^2$}}
\put(423,686){\makebox(0,0)[r]{UFB}}
\put(1031,591){\makebox(0,0)[r]{broken phase}}
\put(1335,209){\makebox(0,0)[r]{unbroken phase}}
\put(220.0,113.0){\rule[-0.200pt]{0.400pt}{184.048pt}}
\put(625,113){\usebox{\plotpoint}}
\multiput(625.00,113.58)(0.539,0.492){21}{\rule{0.533pt}{0.119pt}}
\multiput(625.00,112.17)(11.893,12.000){2}{\rule{0.267pt}{0.400pt}}
\multiput(638.00,125.58)(0.543,0.492){19}{\rule{0.536pt}{0.118pt}}
\multiput(638.00,124.17)(10.887,11.000){2}{\rule{0.268pt}{0.400pt}}
\multiput(650.00,136.58)(0.496,0.492){21}{\rule{0.500pt}{0.119pt}}
\multiput(650.00,135.17)(10.962,12.000){2}{\rule{0.250pt}{0.400pt}}
\multiput(662.00,148.58)(0.543,0.492){19}{\rule{0.536pt}{0.118pt}}
\multiput(662.00,147.17)(10.887,11.000){2}{\rule{0.268pt}{0.400pt}}
\multiput(674.00,159.58)(0.539,0.492){21}{\rule{0.533pt}{0.119pt}}
\multiput(674.00,158.17)(11.893,12.000){2}{\rule{0.267pt}{0.400pt}}
\multiput(687.00,171.58)(0.543,0.492){19}{\rule{0.536pt}{0.118pt}}
\multiput(687.00,170.17)(10.887,11.000){2}{\rule{0.268pt}{0.400pt}}
\multiput(699.00,182.58)(0.496,0.492){21}{\rule{0.500pt}{0.119pt}}
\multiput(699.00,181.17)(10.962,12.000){2}{\rule{0.250pt}{0.400pt}}
\multiput(711.00,194.58)(0.539,0.492){21}{\rule{0.533pt}{0.119pt}}
\multiput(711.00,193.17)(11.893,12.000){2}{\rule{0.267pt}{0.400pt}}
\multiput(724.00,206.58)(0.543,0.492){19}{\rule{0.536pt}{0.118pt}}
\multiput(724.00,205.17)(10.887,11.000){2}{\rule{0.268pt}{0.400pt}}
\multiput(736.00,217.58)(0.496,0.492){21}{\rule{0.500pt}{0.119pt}}
\multiput(736.00,216.17)(10.962,12.000){2}{\rule{0.250pt}{0.400pt}}
\multiput(748.00,229.58)(0.543,0.492){19}{\rule{0.536pt}{0.118pt}}
\multiput(748.00,228.17)(10.887,11.000){2}{\rule{0.268pt}{0.400pt}}
\multiput(760.00,240.58)(0.539,0.492){21}{\rule{0.533pt}{0.119pt}}
\multiput(760.00,239.17)(11.893,12.000){2}{\rule{0.267pt}{0.400pt}}
\multiput(773.00,252.58)(0.543,0.492){19}{\rule{0.536pt}{0.118pt}}
\multiput(773.00,251.17)(10.887,11.000){2}{\rule{0.268pt}{0.400pt}}
\multiput(785.00,263.58)(0.496,0.492){21}{\rule{0.500pt}{0.119pt}}
\multiput(785.00,262.17)(10.962,12.000){2}{\rule{0.250pt}{0.400pt}}
\multiput(797.00,275.58)(0.539,0.492){21}{\rule{0.533pt}{0.119pt}}
\multiput(797.00,274.17)(11.893,12.000){2}{\rule{0.267pt}{0.400pt}}
\multiput(810.00,287.58)(0.543,0.492){19}{\rule{0.536pt}{0.118pt}}
\multiput(810.00,286.17)(10.887,11.000){2}{\rule{0.268pt}{0.400pt}}
\multiput(822.00,298.58)(0.496,0.492){21}{\rule{0.500pt}{0.119pt}}
\multiput(822.00,297.17)(10.962,12.000){2}{\rule{0.250pt}{0.400pt}}
\multiput(834.00,310.58)(0.543,0.492){19}{\rule{0.536pt}{0.118pt}}
\multiput(834.00,309.17)(10.887,11.000){2}{\rule{0.268pt}{0.400pt}}
\multiput(846.00,321.58)(0.539,0.492){21}{\rule{0.533pt}{0.119pt}}
\multiput(846.00,320.17)(11.893,12.000){2}{\rule{0.267pt}{0.400pt}}
\multiput(859.00,333.58)(0.496,0.492){21}{\rule{0.500pt}{0.119pt}}
\multiput(859.00,332.17)(10.962,12.000){2}{\rule{0.250pt}{0.400pt}}
\multiput(871.00,345.58)(0.543,0.492){19}{\rule{0.536pt}{0.118pt}}
\multiput(871.00,344.17)(10.887,11.000){2}{\rule{0.268pt}{0.400pt}}
\multiput(883.00,356.58)(0.539,0.492){21}{\rule{0.533pt}{0.119pt}}
\multiput(883.00,355.17)(11.893,12.000){2}{\rule{0.267pt}{0.400pt}}
\multiput(896.00,368.58)(0.543,0.492){19}{\rule{0.536pt}{0.118pt}}
\multiput(896.00,367.17)(10.887,11.000){2}{\rule{0.268pt}{0.400pt}}
\multiput(908.00,379.58)(0.496,0.492){21}{\rule{0.500pt}{0.119pt}}
\multiput(908.00,378.17)(10.962,12.000){2}{\rule{0.250pt}{0.400pt}}
\multiput(920.00,391.58)(0.543,0.492){19}{\rule{0.536pt}{0.118pt}}
\multiput(920.00,390.17)(10.887,11.000){2}{\rule{0.268pt}{0.400pt}}
\multiput(932.00,402.58)(0.539,0.492){21}{\rule{0.533pt}{0.119pt}}
\multiput(932.00,401.17)(11.893,12.000){2}{\rule{0.267pt}{0.400pt}}
\multiput(945.00,414.58)(0.496,0.492){21}{\rule{0.500pt}{0.119pt}}
\multiput(945.00,413.17)(10.962,12.000){2}{\rule{0.250pt}{0.400pt}}
\multiput(957.00,426.58)(0.543,0.492){19}{\rule{0.536pt}{0.118pt}}
\multiput(957.00,425.17)(10.887,11.000){2}{\rule{0.268pt}{0.400pt}}
\multiput(969.00,437.58)(0.539,0.492){21}{\rule{0.533pt}{0.119pt}}
\multiput(969.00,436.17)(11.893,12.000){2}{\rule{0.267pt}{0.400pt}}
\multiput(982.00,449.58)(0.543,0.492){19}{\rule{0.536pt}{0.118pt}}
\multiput(982.00,448.17)(10.887,11.000){2}{\rule{0.268pt}{0.400pt}}
\multiput(994.00,460.58)(0.496,0.492){21}{\rule{0.500pt}{0.119pt}}
\multiput(994.00,459.17)(10.962,12.000){2}{\rule{0.250pt}{0.400pt}}
\multiput(1006.00,472.58)(0.543,0.492){19}{\rule{0.536pt}{0.118pt}}
\multiput(1006.00,471.17)(10.887,11.000){2}{\rule{0.268pt}{0.400pt}}
\multiput(1018.00,483.58)(0.539,0.492){21}{\rule{0.533pt}{0.119pt}}
\multiput(1018.00,482.17)(11.893,12.000){2}{\rule{0.267pt}{0.400pt}}
\multiput(1031.00,495.58)(0.496,0.492){21}{\rule{0.500pt}{0.119pt}}
\multiput(1031.00,494.17)(10.962,12.000){2}{\rule{0.250pt}{0.400pt}}
\multiput(1043.00,507.58)(0.543,0.492){19}{\rule{0.536pt}{0.118pt}}
\multiput(1043.00,506.17)(10.887,11.000){2}{\rule{0.268pt}{0.400pt}}
\multiput(1055.00,518.58)(0.539,0.492){21}{\rule{0.533pt}{0.119pt}}
\multiput(1055.00,517.17)(11.893,12.000){2}{\rule{0.267pt}{0.400pt}}
\multiput(1068.00,530.58)(0.543,0.492){19}{\rule{0.536pt}{0.118pt}}
\multiput(1068.00,529.17)(10.887,11.000){2}{\rule{0.268pt}{0.400pt}}
\multiput(1080.00,541.58)(0.496,0.492){21}{\rule{0.500pt}{0.119pt}}
\multiput(1080.00,540.17)(10.962,12.000){2}{\rule{0.250pt}{0.400pt}}
\multiput(1092.00,553.58)(0.543,0.492){19}{\rule{0.536pt}{0.118pt}}
\multiput(1092.00,552.17)(10.887,11.000){2}{\rule{0.268pt}{0.400pt}}
\multiput(1104.00,564.58)(0.539,0.492){21}{\rule{0.533pt}{0.119pt}}
\multiput(1104.00,563.17)(11.893,12.000){2}{\rule{0.267pt}{0.400pt}}
\multiput(1117.00,576.58)(0.496,0.492){21}{\rule{0.500pt}{0.119pt}}
\multiput(1117.00,575.17)(10.962,12.000){2}{\rule{0.250pt}{0.400pt}}
\multiput(1129.00,588.58)(0.543,0.492){19}{\rule{0.536pt}{0.118pt}}
\multiput(1129.00,587.17)(10.887,11.000){2}{\rule{0.268pt}{0.400pt}}
\multiput(1141.00,599.58)(0.496,0.492){21}{\rule{0.500pt}{0.119pt}}
\multiput(1141.00,598.17)(10.962,12.000){2}{\rule{0.250pt}{0.400pt}}
\multiput(1153.00,611.58)(0.590,0.492){19}{\rule{0.573pt}{0.118pt}}
\multiput(1153.00,610.17)(11.811,11.000){2}{\rule{0.286pt}{0.400pt}}
\multiput(1166.00,622.58)(0.496,0.492){21}{\rule{0.500pt}{0.119pt}}
\multiput(1166.00,621.17)(10.962,12.000){2}{\rule{0.250pt}{0.400pt}}
\multiput(1178.00,634.58)(0.543,0.492){19}{\rule{0.536pt}{0.118pt}}
\multiput(1178.00,633.17)(10.887,11.000){2}{\rule{0.268pt}{0.400pt}}
\multiput(1190.00,645.58)(0.539,0.492){21}{\rule{0.533pt}{0.119pt}}
\multiput(1190.00,644.17)(11.893,12.000){2}{\rule{0.267pt}{0.400pt}}
\multiput(1203.00,657.58)(0.496,0.492){21}{\rule{0.500pt}{0.119pt}}
\multiput(1203.00,656.17)(10.962,12.000){2}{\rule{0.250pt}{0.400pt}}
\multiput(1215.00,669.58)(0.543,0.492){19}{\rule{0.536pt}{0.118pt}}
\multiput(1215.00,668.17)(10.887,11.000){2}{\rule{0.268pt}{0.400pt}}
\multiput(1227.00,680.58)(0.496,0.492){21}{\rule{0.500pt}{0.119pt}}
\multiput(1227.00,679.17)(10.962,12.000){2}{\rule{0.250pt}{0.400pt}}
\multiput(1239.00,692.58)(0.590,0.492){19}{\rule{0.573pt}{0.118pt}}
\multiput(1239.00,691.17)(11.811,11.000){2}{\rule{0.286pt}{0.400pt}}
\multiput(1252.00,703.58)(0.496,0.492){21}{\rule{0.500pt}{0.119pt}}
\multiput(1252.00,702.17)(10.962,12.000){2}{\rule{0.250pt}{0.400pt}}
\multiput(1264.00,715.58)(0.496,0.492){21}{\rule{0.500pt}{0.119pt}}
\multiput(1264.00,714.17)(10.962,12.000){2}{\rule{0.250pt}{0.400pt}}
\multiput(1276.00,727.58)(0.590,0.492){19}{\rule{0.573pt}{0.118pt}}
\multiput(1276.00,726.17)(11.811,11.000){2}{\rule{0.286pt}{0.400pt}}
\multiput(1289.00,738.58)(0.496,0.492){21}{\rule{0.500pt}{0.119pt}}
\multiput(1289.00,737.17)(10.962,12.000){2}{\rule{0.250pt}{0.400pt}}
\multiput(1301.00,750.58)(0.543,0.492){19}{\rule{0.536pt}{0.118pt}}
\multiput(1301.00,749.17)(10.887,11.000){2}{\rule{0.268pt}{0.400pt}}
\multiput(1313.00,761.58)(0.496,0.492){21}{\rule{0.500pt}{0.119pt}}
\multiput(1313.00,760.17)(10.962,12.000){2}{\rule{0.250pt}{0.400pt}}
\multiput(1325.00,773.58)(0.590,0.492){19}{\rule{0.573pt}{0.118pt}}
\multiput(1325.00,772.17)(11.811,11.000){2}{\rule{0.286pt}{0.400pt}}
\multiput(1338.00,784.58)(0.496,0.492){21}{\rule{0.500pt}{0.119pt}}
\multiput(1338.00,783.17)(10.962,12.000){2}{\rule{0.250pt}{0.400pt}}
\multiput(1350.00,796.58)(0.496,0.492){21}{\rule{0.500pt}{0.119pt}}
\multiput(1350.00,795.17)(10.962,12.000){2}{\rule{0.250pt}{0.400pt}}
\multiput(1362.00,808.58)(0.590,0.492){19}{\rule{0.573pt}{0.118pt}}
\multiput(1362.00,807.17)(11.811,11.000){2}{\rule{0.286pt}{0.400pt}}
\multiput(1375.00,819.58)(0.496,0.492){21}{\rule{0.500pt}{0.119pt}}
\multiput(1375.00,818.17)(10.962,12.000){2}{\rule{0.250pt}{0.400pt}}
\multiput(1387.00,831.58)(0.543,0.492){19}{\rule{0.536pt}{0.118pt}}
\multiput(1387.00,830.17)(10.887,11.000){2}{\rule{0.268pt}{0.400pt}}
\multiput(1399.00,842.58)(0.496,0.492){21}{\rule{0.500pt}{0.119pt}}
\multiput(1399.00,841.17)(10.962,12.000){2}{\rule{0.250pt}{0.400pt}}
\multiput(1411.00,854.58)(0.590,0.492){19}{\rule{0.573pt}{0.118pt}}
\multiput(1411.00,853.17)(11.811,11.000){2}{\rule{0.286pt}{0.400pt}}
\multiput(1424.00,865.58)(0.496,0.492){21}{\rule{0.500pt}{0.119pt}}
\multiput(1424.00,864.17)(10.962,12.000){2}{\rule{0.250pt}{0.400pt}}
\end{picture}

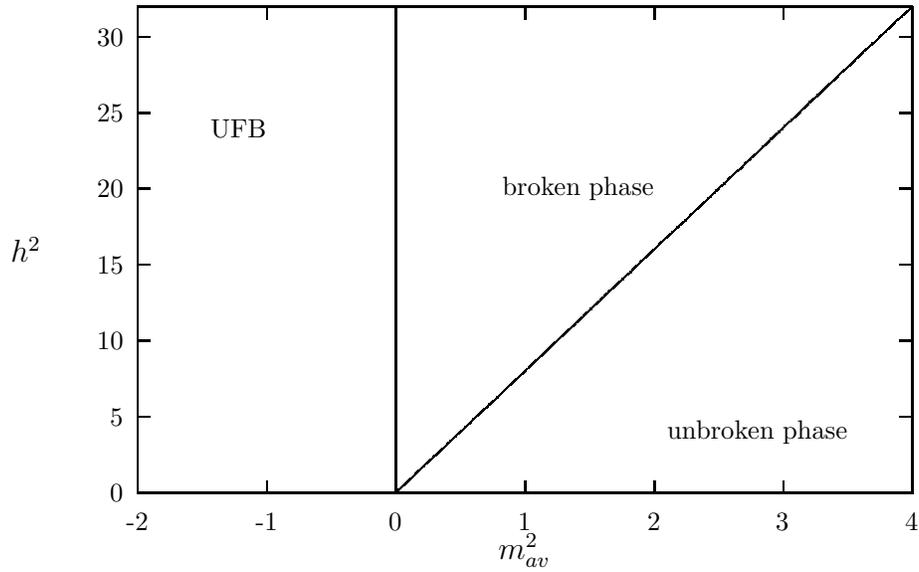
Fig. 2: The phase structure for $\rho =1$.
UFB denotes the unbounded-from-below direction.
\end{center}
\begin{center}
\input{fig3.tex}

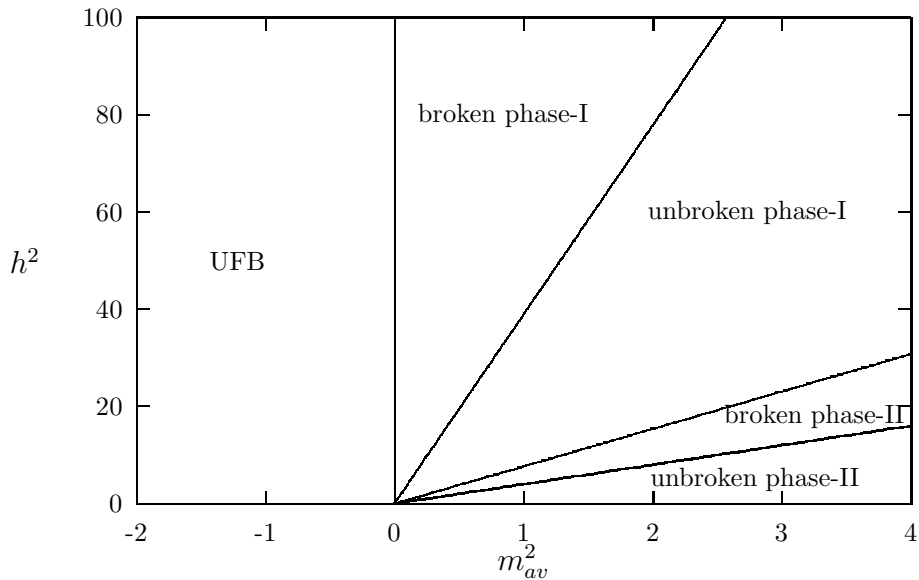
Fig. 3: The phase structure for $\rho =20$.
\end{center}

In the unbroken phase none of local or global symmetries is broken 
except the $R$-symmetry.
All the dual quarks and singlet fermions  $\chi_T$ are massless, 
although their scalar partners become massive 
due to the soft mass terms.
In the broken phase the dual gauge symmetry is broken completely and 
$\tilde N_c$ flavours of dual quarks become massive .
Only $N_c$ ($=N_f-\tilde N_c$) flavour of dual quarks and $N_c \times N_c$ 
singlet fermions $\chi_T$ remain massless \footnote{
Here the term dual ``quark'' has not to be taken literally since  
the dual gauge group is completely broken.
Thus, these dual quarks are nothing but singlet fermions.}.
They have the global symmetry 
$SU(N_c)_q\times SU(N_c)_{\bar q} \times U(1)_{B'}$.
All the scalar fields become massive.
This breaking pattern is similar to the one discussed 
in Ref. \cite{soft2}, but slightly different.
In Ref. \cite{soft2}, the flavour symmetry is broken by hand, i.e.  
by nondegenerate soft scalar masses, 
while some of them are taken to be imaginary.
That leads to the same type of gauge symmetry breakdown.
However, in that model the fields $T$ do not develop their 
vacuum expectation values.
On the other hand, in our model sponteneous symmetry breaking 
can occur without the breaking of flavour symmetry by hand even for 
positive values of soft scalar mass squared.
In addition, the fields $T$ also develop their 
vacuum expectation values in the broken phase of our model.

\section{Softly broken dual pair}
In this section we consider the relation between softly broken 
original SUSY QCD and dual theories.
In the original theory the soft scalar mass terms as well as gaugino 
mass terms are all we can add as soft SUSY breaking terms, i.e. 
\begin{equation}
{\cal L_{SB}}=-m_Q^2|Q|^2-m_{\bar Q}^2|\bar Q|^2.
\end{equation}

Let us discuss the unbroken phase $q=\bar q=T=0$, where one of 
the conditions (\ref{cond1}) and (\ref{cond2}) is not satisfied.
In this case the structure of massless fermions and global symmetries 
except gauginos and $R$-symmetry is not changed compared with the SUSY 
limit.
Thus, this case leads to the same anomaly structure 
for the unbroken global symmetry 
$SU(N_f)_q \times SU(N_f)_{\bar q} \times U(1)_B$ 
as the SUSY limit.
On the other hand, we have the unbroken phase $Q=\bar Q=0$ for 
$m_Q^2 > 0$ and $m_{\bar Q}^2 > 0$.
In this case no local or global symmetry is broken except 
the $R$-symmetry, which is broken by gaugino mass terms.
Moreover, all the quarks remain massless.
Thus the anomaly structure is the same as for the SUSY limit, e.g. 
\begin{equation}
SU(N_f)^3 \quad {\rm and} \quad SU(N_f)^2U(1)_B.
\end{equation}
Therefore, this dual pair has the same anomaly structure in the 
unbroken phase even in the presence of soft SUSY breaking terms.
That seems to imply the presence of Seiberg's duality in this phase 
even after SUSY breaking with the $A$-terms.
This observation has been already made in Ref. \cite{soft1}, although 
the $A$-terms were not included in the discussions.

Let us extend the above consideration to the broken phase and notice 
that large symmetry breaking takes place in the broken phase 
discussed in the previous section.
Here we simplify the issue and consider the following model.
When adding the soft SUSY breaking terms, we break the flavour 
symmtry $SU(\tilde N_c)_q \times SU(\tilde N_c)_{\bar q}$ into 
$SU(\tilde N_c-1)_q \times U(1)_q \times 
SU(\tilde N_c-1)_{\bar q} \times U(1)_{\bar q}$.
Then we assume the first flavour has soft scalar masses,   
$m_{q1}$ and $m_{\bar q 1}$, different from the others, 
$m_q$ and $m_{\bar q}$ \cite{soft2} \footnote{
This scenario where only one flavour is different from the others, could 
be conceivable in the same way as the top quark is much heavier 
than the rest in the real world.}.
Recall that the $i$-th flavour is decoupled from the other 
flavours in all the conditions and equations to realize the broken phase.
Here we assum that only the soft scalar masses of the first flavour, 
$m_{q1}$ and $m_{\bar q 1}$, satisfy the breaking conditions, 
(\ref{cond1}) and (\ref{cond2}).
In this case only the vacuum expectation values $X_1$ and $T_{(1)}$  
are developed.
That leads to the gauge symmetry breaking, 
\begin{equation}
SU(\tilde N_c) \rightarrow SU(\tilde N_c -1)\  (=SU(N_f-1-N_c)).
\end{equation}
Furthermore, $(N_f-1)$ flavours of dual quarks and $(N_f-1) \times (N_f-1)$ 
singlet fermions $\chi_T$ remain massless.
These massless fermions have the global symmetry 
$SU(N_f-1)_q \times SU(N_f-1)_{\bar q} \times U(1)_{B'}$.
Massless dual quarks, $\psi_q$ and $\psi_{\bar q}$, and singlet 
fermions $\chi_T$ transform as $(\bar N_f,0,N_c/(\tilde N_c -1))$, 
$(0,N_f,-N_c/(\tilde N_c -1))$ and $(N_f,\bar N_f,0)$ under this 
global symmetry, respectively.
All the scalar fields become massive.
This structure of massless fermions obviously corresponds to the 
SUSY model with $SU(N_f-1-N_c)$ gauge group and $(N_f-1)$ flavours of quarks.
This SUSY model is dual to SUSY QCD theory with $SU(N_c)$ gauge 
group and $(N_f-1)$ flavours of quarks.

Let us consider now the corresponding original theory.
If at the SUSY breaking scale, the  
flavour symmetry is broken in the same way as the one of the dual theory, 
$SU(N_f-1)_q \times SU(N_f-1)_{\bar q} $, nothing would prevent the  
appearance of the following superpotential: 
\begin{equation}
W=M_1\widehat Q^1 \widehat {\bar Q_1}.
\label{mmixing}
\end{equation} 
Note that in this case the $B$-term, $-M_B^2Q^1\bar Q_1$, can 
also appear as the soft terms in the lagrangian ${\cal L_{SB}}$.
Thus, the (mass)$^2$ matrix of the first flavour of squarks, $M_{11}^2$ 
is written as 
\begin{equation}
M_{11}^2 = \left( \begin{array}{cc}
m_{Q1}^2+M_1^2 & -M_B^2 \\
-M_B^2 & m_{\bar Q1}^2 +M_1^2
   \end{array}\right).
\label{Mmatrix}
\end{equation}
If $\det(M_{11}^2) > 0$, the potential minimum corresponds to 
$Q^1=\bar Q_1=0$ and 
the gauge symmetry $SU(N_c)$ remains unbroken.
In this case $(N_f-1)$ flavours of quarks remain massless and 
these massless fermions have the global symmetry  
$SU(N_f-1)_q \times SU(N_f-1)_{\bar q}\times U(1)_B$.
All scalar fields become massive.
This model has the same anomaly structure as the softly 
broken dual thery in the broken phase, e.g. for 
\begin{equation}
SU(N_f-1)^3 \quad {\rm and} \quad SU(N_f-1)^2U(1)_B,
\end{equation}
where $U(1)_B$ should be replaced by $U(1)_{B'}$ in the dual theory.
That seems to suggest the presence of Seiberg's duality after SUSY breaking 
even in the broken phase.

Let us discuss the case with $\det(M_{11}^2) < 0$ and  
$m_{Q1}^2+m_{\bar Q1}^2 + 2M_1^2 >2|M_B^2|$.
In this case squarks $Q^1$ and $\bar Q_1$ develop their 
finite vacuum expectation values and the gauge symmetry is broken into 
$SU(N_c-1)$.
Only the $(N_f-1)$ flavours of quarks remain massless and they have 
the global symmetry 
$SU(N_f-1)_Q \times SU(N_f-1)_{\bar Q} \times U(1)_B$.
This case seems to correspond to the dual theory for the 
unbounded-from-below direction, i.e. $m_{T1}^2 < 0$, where 
the $SU(\tilde N_c)$ gauge symmetry is unbroken and 
$(N_f-1)$ flavours of dual quarks remain massless.
We note that such a dual theory along this specific direction 
has no stable vacuum within its own framework, 
and thus this unbounded-from-direction can not be described 
within the framewoek of the dual theory.

We have considered the case where only one flavour of squarks develop 
their vacuum expectation values.
We can easily extend the above discussion to the case when 
more flavours of squarks develop their vacuum expectaion values.
Then we can obtain similar relations between softly broken 
orginal and dual theories in the broken phase.

Let us also give some comments on the unbounded-from-below directions 
for $m_q^2 <0$.
In this case the scalar potential of the dual theory is 
unbounded from below and vacuum expectation values of 
$q_{(i)}$ and the baryonic operator $b=\prod q_{(i)}$ 
run away to infinity, $q_{(i)} \rightarrow \infty$.
This baryonic operator corresponds to the baryonic operator of 
$Q_i$ in the original theory.
Thus the unbounded-from-below direction for $m_q^2 <0$ in the dual theory 
corresponds to the unbounded-from-below direction for 
$m_Q^2 <0$, where vacuum expectation values of $Q_i$ and their 
baryonic operator go to infinity.
We have the same situation for $m_{\bar q}^2 <0$ and 
$m_{\bar Q}^2 <0$.

\section{Conclusions}

We have studied the softly broken SUSY QCD taking 
into account the effects of $A$-terms.
We have investigated the phase structure of the softly broken 
dual theory and 
have found that the $A$-terms play a basic role 
in the realization of the broken phase.
Also we have found relations between softly broken dual 
pair even in the broken phase.
These results should be useful in the understanding of QCD and confinment 
in the real world.
Detailed quantiative results, including the mass spectra, will be 
discussed elsewhere.

Seiberg's duality can be understood from the viewpoint of $D$-brane 
dynamics.
It is interesting to study our results in non-SUSY cases also from 
the viewpoint of $D$-brane dynamics.

One could discuss in a similar way the case with $N_f \leq N_c+1$ as 
treated in Ref.\cite{soft1} but with generic soft breaking terms 
in order to investigate whether they play any specific role.

\section*{Acknowledgments}  
This work was partially supported by the Academy of Finland under 
Project no. 37599.


\begin{thebibliography}{99}

\bibitem{rev}
For reviews see, e.g.
K.~Intriligator and N.¨Seiberg, 
Nucl. Phys. Proc. Suppl. {\bf 45BC} (1996) 1;\\
L.~\'Alvarez-Gaum\'e and S.F.~Hassan, CERN-TH/96-371, hep-th/9701069;\\
M.~Shifman, TPI-MINN-97/09-T, hep-th/9704114.


\bibitem{seiberg1}
N. Seiberg, Phys. Rev. {\bf D49} (1994) 6857;\\
K. Intriligator, R.G. Leigh and N. Seiberg, 
Phys. Rev. {\bf D49} (1994) 1092;\\
N. Seiberg, {\it The Power of Holomorphy -- Exact Results in 
4D SUSY Field Theory}, talk given at the PASCOS 94, {\tt hep-th/9408013}.

\bibitem{MO}
C. Montonen and D. Olive, Phys. Lett. {\bf B72} (1977) 117.

\bibitem{osborn}
H. Osborn, Phys. Lett. {\bf B83} (1979) 321.

\bibitem{SW}
N. Seiberg and E. Witten, Nucl. Phys. {\bf B426} (1994) 19;
(E) {\bf B430} (1994) 485;\\
N. Seiberg and E. Witten, Nucl. Phys. {\bf B431} (1994) 484.

\bibitem{seiberg2}
N. Seiberg, Nucl. Phys. {\bf B435} (1995) 129.

\bibitem{thooft}
G. 't Hooft, in {\it Recent Developments in Gauge Theories},
 eds. G. 't Hooft et al., 135 (Plenum Press, New York, 1980).

\bibitem{soft1}
O.~Aharony, J.~Sonnenschein, M.E.~Peskin and S.~Yankielowicz, 
Phys. Rev. {\bf D52} (1995) 6157.

\bibitem{soft2}
E.~D'Hoker, Y.~Mimura and N.~Sakai, 
Phys. Rev. {\bf D54} (1996) 7724.

\bibitem{soft3}
N.~Evans, S.D.H.~Hsu and M.~Schwets, 
Phys. Lett. {\bf B404} (1997) 77;\\
H.C.~Cheng and Y.~Shadmi, Fermilab-Pub-97/420-T,
hep-th/9801146;\\
S.P.~Martin and J.~D.~Wells, SLAC-PUB-7739, hep-th/9801157.

\bibitem{soft4}
N.~Evans, S.D.H.~Hsu, M.~Schwets and S.B.~Selipsky,
Nucl. Phys. {\bf B456} (1995) 205;\\
N.~Evans, S.D.H.~Hsu and M.~Schwets, 
Nucl. Phys. {\bf B484} (1997) 124;\\
L.~\'Alvarez-Gaum\'e, J.~Distler, C.~Kounnas and M.~Mari\~no,
Int. J. Mod. Phys. {\bf 11} (1996) 4745;\\
L.~\'Alvarez-Gaum\'e and M.~Mari\~no,
Int. J. Mod. Phys. {\bf 12} (1997) 975;\\
K.~Konishi, Phys. Lett. {\bf B392} (1997) 101;\\
L.~\'Alvarez-Gaum\'e, M.~Mari\~no and F.~Zamora,
CERN-TH/97-37, hep-th/9703072; CERN-TH/97-144, hep-th/9707017.

\bibitem{Higgs}
For a review, see L.E.~Ib\'a\~nez and G.G.~Ross, in 
{\it Perspectives in Higgs Physics}, ed. G.~Kane (World Scientfic, 1992).

\bibitem{CCB}
J.M.~Fr\`ere, D.R.T.~Jones and S.~Raby, Nucl. Phys. {\bf B222} (1983) 11;\\
L.~Alvarez-Gaum\'e, J.~Polchinski and M.~Wise, 
Nucl. Phys. {\bf B221} (1983) 495;\\
J.P.~Derendinger and C.A.~Savoy, Nucl. Phys. {\bf B237} (1984) 307;\\
C.~Kounnas, A.B.~Lahanas, D.V.~Nanopoulos and M.~Quir\'os,
Nucl. Phys. {\bf B236} (1984) 438.

\bibitem{ST-soft}
L.E.~Ib\'a\~nez and D.~L\"ust, Nucl. Phys. {\bf B382} (1992) 305;\\
V.S.~Kaplunovsky and J.~Louis, Phys. Lett. {\bf B306} (1993) 269;\\
A.~Brignole, L.E.~Ib\'a\~nez and C.~Mu\~noz,
 Nucl. Phys. {\bf B422} (1994) 125.

\bibitem{dsb}
See, e.g. G.F.~Giudice and R.~Rattazzi, CERN-TH-97-380, hep-ph/9801271, 
and references therein.

\bibitem{cred}
W.~Zimmermann, Com. Math. Phys. {\bf 97} (1985) 211;\\
R.~Oehme and W.~Zimmermann, Com. Math. Phys. {\bf 97} (1985) 569.

\bibitem{RGsoft}
Y.~Kawamura, T.~Kobayashi and J.~Kubo, 
Phys. Lett. {\bf B405} (1997) 64;\\
T.~Kobayashi, J.~Kubo, M.~Mondrag\'on and G.~Zoupanos, 
Nucl. Phys. {\bf B511} (1998) 45;\\
T.~Kobayashi, J.~Kubo and G.~Zoupanos, 
HIP-1998-06/TH, hep-ph/9802267.


\end{thebibliography}
\end{document}